\documentclass[aps,pre,twocolumn,superscriptaddress,floatfix]{revtex4-1}
\usepackage{graphicx,amsmath,amssymb,subfigure,epsfig,psfrag,verbatim,color}
\begin{document}
\title{Connecting Complex Electronic Pattern Formation to Critical Exponents}

\author{Shuo Liu}
\email{liu305@purdue.edu} \affiliation{Department of Physics, Purdue
University, West Lafayette, IN 47907, USA}
\author{E.~W. Carlson}
\affiliation{Department of Physics, Purdue University, West
Lafayette, IN 47907, USA}
\author{K.~A. Dahmen}
\affiliation{Department of Physics, University of Illinois,
Urbana-Champaign, IL 61801, USA}
\date{\today}

\begin{abstract}
Scanning probes reveal complex, inhomogeneous patterns on the surface
of many condensed matter systems. In some cases, the
patterns form self-similar, fractal geometric clusters. In this paper,
we advance the theory of criticality as it pertains to those geometric
clusters (defined as connected sets of nearest-neighbor aligned spins)
in the context of Ising models.
We show how data from surface probes can be used to distinguish
whether electronic patterns observed at the surface of a material
are confined to the surface, or whether the patterns originate in the bulk.
Whereas thermodynamic critical exponents are derived from
the behavior of Fortuin-Kasteleyn (FK) clusters, critical exponents
can be similarly defined for geometric clusters.
We find that these geometric critical exponents are not only distinct numerically from the thermodynamic and uncorrelated percolation exponents, but that they separately satisfy scaling relations at the critical fixed points  discussed in the text.
We furthermore find that the two-dimensional (2D) cross-sections of geometric
clusters
in the three-dimensional  (3D) Ising model display critical scaling
behavior at the bulk phase transition temperature.
In particular, we show that when considered on a 2D slice
of a 3D system, the pair connectivity function familiar
from percolation theory displays more robust critical behavior than
the spin-spin correlation function, and we calculate the corresponding critical exponent.
We discuss the implications of these two distinct length scales in Ising models.
We also calculate the pair connectivity exponent in the clean 2D case.
These results extend the theory of geometric criticality in the
clean Ising universality classes, and facilitate the broad application of geometric cluster analysis techniques to maximize the information that
can be extracted from scanning image probe data in
condensed matter systems.

\end{abstract}

\maketitle

\section{Introduction}
\label{sxn:introduction}

Since their invention in 1982, scanning probes have revolutionized
our understanding of materials and their surfaces, yielding an ever
increasing wealth of data available for in-depth analysis, on a wide
variety of systems \cite{stm-rmp-nobel}. Theory has been sprinting
to catch up in order to interpret it all, and also to discover new
ways of extracting information from the data, in order to fully
realize this promise of new knowledge from the increasing variety of
scanning probes and their ever-increasing experimental capabilities.
To date, the majority of theoretical treatments have focused on
microscopic physics \cite{basov-rmp}, with few theoretical
treatments offering guidance for how to interpret the wealth of
information available in the  multiscale pattern formation often
observed on surfaces.  We have recently pioneered a new set of
techniques for analyzing scanning probe data by mapping
two-component image data to random Ising models
\cite{phillabaum-2012}, based on geometric cluster methods imported
from disordered statistical mechanics. A geometric cluster is
defined as a set of aligned nearest neighbor sites. The key insight
is that near criticality, the spatial configurations of geometric
clusters are controlled by the critical fixed point, and therefore
the geometric properties encode critical exponents. The method is
capable of extracting information from the data about disorder,
interactions, and dimension.  We have already successfully applied
this new technique to uncover a unification of the fundamental
physics governing the multiscale pattern formation observed in two
disparate strongly correlated electronic materials (cuprate
superconductors \cite{phillabaum-2012,superstripes-erice-2014} and
vanadium dioxide \cite{shuo-vo2}).

However, because it is the Fortuin-Kasteleyn (FK) clusters
\footnote{Fortuin-Kasteleyn clusters are generated from the
geometric clusters by deleting bonds of a geometric cluster with a
Boltzmann-like probability \cite{fortuin}} which encode
thermodynamic criticality \cite{Hu1984}, rather than the geometric
clusters which are directly accessible experimentally via scanning
probes and which we employ in our method, little is known about the
general theoretical structure of geometric clusters in random Ising
models, and the critical exponents associated with the geometric
clusters are unknown for many of the fixed points which are key to
interpreting experimental data. In this paper, we advance the theory
of criticality as it pertains to geometric clusters in clean Ising
models to further develop geometric cluster analysis techniques
\cite{phillabaum-2012}, in order to maximize the information that
can be extracted from experiments using these new methods.
Although the geometric clusters
do not encode thermodynamic criticality,
we conjecture that when the geometric clusters percolate, whether at
or below the thermodynamic critical temperature, the geometric
clusters do encode {\bf geometric criticality}, complete with its
own set of critical exponents, which we further conjecture are
distinct from the exponents of uncorrelated percolation when arising
in the context of an interacting model.

This paper is organized as follows.
We first describe in Sec.~\ref{sxn:model} the model under consideration.
We then ask in Sec.~\ref{sxn:cxn}
 whether the pair connectivity function can be a power law in any system
other than uncorrelated percolation (to which we will answer
``yes'').   In Sec.~\ref{sxn:exponents}, we present some conjectures
considering geometric criticality as distinct from thermodynamic
criticality. Next, we present in Sec.~\ref{sxn:C-3Dx} our results
for the critical geometric cluster exponents on 2D slices of the
clean 3D Ising model (denoted by C-3Dx, where x means
cross-section), and in Sec.~\ref{sxn:C-2D} our simulations of the
pair connectivity function of the clean 2D Ising model (denoted by
C-2D),
In Sec.~\ref{sxn:coniglio} we discuss the relation of the order
parameter in a 3D ferromagnetic Ising model to the percolation of
geometric clusters on {\em 2D slices} of the 3D system and derive
that the position of the 2D slice geometric cluster percolation
point coincides with bulk critical point. In Sec.~\ref{sxn:cl} we
discuss the implications of the pair connectivity length scale in
Ising models. Finally, in Sec.~\ref{sxn:discussion}, we discuss and
summarize our findings about geometric criticality, and present
conclusions in Sec.~\ref{sxn:conclusions}.

\section{The Model}
\label{sxn:model}

In strongly correlated electronic systems, the combination of
disorder and strong correlations can drive complex pattern formation
\cite{Dagotto}, but disentangling correlations from disorder in the
experimental system is an open problem.  Recently, we have developed
new cluster analysis techniques for interpreting scanning image
probe data in cases where the spatial data can be abstracted to two
components (and thus may be mapped to an Ising variable), and where
the resulting cluster patterns display structure on multiple length
scales \cite{phillabaum-2012}.

Scanning probe experiments often reveal complex pattern formation at
the surface of strongly correlated electronic systems \cite{Dagotto,
dagotto-moreo-manganite}. For example,  charge stripe orientations
display complex geometric patterns at the surface of some cuprate
superconductors, as revealed by scanning tunneling microscopy
\cite{phillabaum-2012,kohsaka-science-2007,hoffman-clusters}.
Complex patterns have also been observed in thin films of VO$_2$ as
it transitions from metal to insulator, via scanning near-field
infrared microscopy \cite{basov-science,shuo-vo2}. Using the cluster
analysis techniques we developed, we showed that in both cases the
dominant type of disorder driving the pattern formation is in the
random field universality class, and we argued that this is the
origin of nonequilibrium behavior in both systems
\cite{phillabaum-2012,superstripes-erice-2014,shuo-vo2}.

In this paper, we expand the diagnostic toolset of the geometric cluster analyses we have developed to include the functional form and qualitative behavior of the connectivity function, derived directly from geometric clusters.  We find that the pair connectivity function can be a power law at more than just uncorrelated percolation points, and we compute the corresponding critical exponent via Monte Carlo simulations.

We consider a general
short-range Ising model:
\begin{eqnarray}
H&=&-\sum_{\langle ij\rangle_\parallel}(J^\parallel +\delta
J_{ij}^\parallel)\sigma_i \sigma_j -\sum_{\langle
ij\rangle_{\perp}}(J^\perp+\delta J_{ij}^\perp)\sigma_i
\sigma_j \nonumber \\
&-&\sum_i(h+h_i)\sigma_i~,
\label{eqn:model}
\end{eqnarray}
where the sum runs over the sites of a cubic lattice, chosen with
spacing at least as small as the resolution of the images to be
studied. The tendency for neighboring regions in the data image to
be of like character is modeled as a nearest neighbor ferromagnetic
interaction. $J^\parallel$ sets the overall strength of the in-plane
ferromagnetic coupling between nearest-neighbor Ising variables, and
$J^\perp$ represents the overall coupling strength between Ising
variables in neighboring planes. The field $h$ represents a
generalized external field which couples with the local Ising
variables. Microscopic models of non-frustrated disorder flow to two
broad classes of disorder under the renormalization group
transformation: random bond disorder in the coupling strength
(through the term $\delta J_{ij}$), also known as random $T_c$
disorder, and random field disorder couplng with the Ising variable
(through the term $h_i$) \cite{cardy-book}.

The Ising variable $\sigma_i=\pm 1$ on each coarse-grained site
represents one of the two possible states, such as the two possible
electron nematic orientations in the cuprates, or the two states of
conductivity (metallic or insulating) in VO$_2$. For the two
examples above, our cluster analysis
\cite{phillabaum-2012,superstripes-erice-2014,shuo-vo2} of the image
data shows that the geometric clusters (which are defined as the
connected set of the nearest neighbor sites with Ising variables
being the same value) extracted from the multi-scale pattern
formation display universal scaling behavior over multiple decades,
suggesting criticality and universality as the origin of  the
spatial complexity revealed by scanning probe microscopy in strongly
correlated electronic systems. By comparing the data-extracted
critical exponents derived from the self-similarity of the geometric
clusters with the theoretical values for the fixed points contained
in Eqn.~\ref{eqn:model}, we have shown \cite{phillabaum-2012} that
it is possible to identify the universality class governing the
scaling behavior of the geometric clusters observed on surfaces of
novel materials.

Once the universality class driving the pattern formation has been
identified, this in turn yields information about the relative
importance of disorder and interactions,  the dominant type of
disorder in the system, and the dimension of the phenomenon studied
\cite{hoffman-clusters,phillabaum-2012,superstripes-erice-2014,shuo-vo2}.
For example, by extracting the critical exponents associated with
the geometric clusters appearing at the surface of a material, it is
possible to understand whether those clusters are forming merely due
to surface effects, or whether the clusters form throughout the bulk
of the material and then intersect the surface, because the critical
exponents are sensitive to dimension.

\section{Where is the connectivity function power law?}
\label{sxn:cxn}

Although geometric clusters do not encode thermodynamic criticality,
we claim that whenever the geometric clusters percolate, the
resulting power law behavior encodes a type of {\em geometric
criticality} \cite{Coniglio1977,dotsenko-1995}.  We further claim
that such geometric critical points in interacting models are new,
distinct fixed points, which are different from uncorrelated
percolation, as supported by our results to be presented in this
paper. The pair connectivity function, derived from geometric
clusters, is known to be  power law at uncorrelated percolation
points. The pair connectivity function $g_{\rm conn}(r)$ is defined
as the probability that two sites separated by a distance $r$ belong
to the same connected finite cluster.
Can it also be power law at certain fixed points in interacting models?
The most natural clusters to define in Ising models are the geometric clusters,
defined by nearest-neighbor sets of like Ising ``spins.''
The geometric clusters have the advantage that they are  directly accessible
in image probe data in a 2-component system, and
they are the clusters we employ in our method.

However,
geometric clusters in Ising models
are still poorly understood theoretically, presumably because they
do not necessarily percolate at the thermodynamic transition
temperature, $T_c$.
Rather, it is the FK clusters
which percolate at the thermodynamic transition temperature $T_c$,
and which encode the thermodynamic critical behavior \cite{Hu1984}.
The FK clusters also constitute the critical ``droplets'' of the
Fisher droplet model \cite{fisher-droplet}. In the case of the clean
2D Ising model (C-2D), it has been shown that the geometric clusters
also percolate at the thermodynamic transition, where $T_p = T_c$
\cite{Coniglio1977}.
In other cases, such as the clean and random bond Ising models in
three dimensions, it has been shown that the bulk geometric clusters
percolate at a temperature $T_p < T_c$, {\em inside the ordered
phase}
\cite{sykes-gaunt-1976,dotsenko-1995,Coniglio1977,berche-janke-2004}.

For the thermodynamic fixed points of Eqn.~\ref{eqn:model} at which
$T_p = T_c$ (as happens at C-2D), the geometric clusters have
well-defined fractal dimensions \cite{janke-clusters-2005}   both
fractal volume dimension $d_v$ and fractal hull dimension $d_h$. In
addition, geometric clusters defined on a {\em 2D slice} of the
clean 3D Ising model also display fractal behavior
\cite{saberi-C3Dx} At these points, because of their fractal
structure, the large scale geometric clusters are self-similar, {\em
i.e.} they look the same on all length scales. This is a tell-tale
characteristic of power law behavior, which brings us leading to our
first conjecture: \vspace{4pt}
\\
{\bf Conjecture \#1:}  The connectivity function is a power law at
all critical points points for which the geometric clusters have
fractal dimensions, $d_v$ and $d_h$.
\vspace{11pt}
\\

Our  results in Sec.~\ref{sxn:C-2D} indicate that indeed the
connectivity function is a power law at the C-2D critical fixed
point, as shown in Figs.~\ref{fig:C-2D-cxn} and
\ref{fig:C-2D-cxn-T}, in support of Conjecture \#1. This is of
course implicit in the pioneering work of Coniglio and coworkers on
the percolation of geometric clusters in this model
\cite{Coniglio:1980uo,Coniglio1977}. Our results in
Sec.~\ref{sxn:C-3Dx} show that the connectivity function is also a
power law {\em on a 2D slice} as the clean 3D Ising system passes
through thermodynamic criticality at $T_c^{3D}$. (See
Figs.~\ref{fig:C-3Dx-cxn} and \ref{fig:C-3Dx-cxn-T}.) It was already
known from the work of Ref.~\cite{saberi-C3Dx} that geometric
clusters defined on a 2D slice have well-defined fractal dimensions
at $T_c^{3D}$ in the clean 3D Ising model, lending further support
to Conjecture \#1.

\section{Geometric Criticality and Thermodynamic Criticality: Two Types of Critical Exponents}
\label{sxn:exponents}

It is known that at the clean 2D Ising (C-2D) fixed point, there are
two distinct sets of critical exponents:  one set associated with
the thermodynamic criticality and encoded by the FK clusters, and a
different set associated with ``geometric criticality'', encoded by
the geometric clusters
\cite{Hu1984,Coniglio1977,janke-clusters-2005}. Because both types
of clusters percolate right at the transition temperature, $T_p =
T_c$, both types of clusters exhibit power law behavior, yielding
two distinct sets of critical exponents, one set derived from each
type of cluster. While the scaling laws governing the exponent
relations at this fixed point are the same for the two types of
clusters, the values of the geometric exponents are not equal to the
values of the thermodynamic exponents, and neither set are equal to
the values of uncorrelated percolation exponents.  In this sense,
the C-2D fixed point is therefore both a thermodynamic critical
point and a type of geometric critical
point.%
This also implies that there are two order parameters for the 2D
clean Ising model (the magnetization and also the infinite cluster);
as we will show below, corresponding to the two order parameters are
two distinct length scales.

The set of critical exponents  derived from the thermodynamic order
parameter (in this case, the magnetization), is associated with
percolation of the FK clusters, and encodes  thermodynamic
criticality -- we will denote these exponents by the subscript ``c''
(to match the ``c'' in the thermodynamic critical temperature
$T_c$.) This type of exponent has been widely discussed in the
literature. The other set of  exponents is derived from the
percolation order parameter, which is the infinite network strength,
defined as the ratio of the number of sites in the infinite
connected geometric cluster to the total number of sites in the
system. This second set of critical exponents is associated with
percolation of geometric clusters, and encodes  geometric
criticality -- we will denote these exponents by the subscript ``p''
(for percolation). The fact that there can be two distinct sets of
exponents has been largely ignored in the literature, except at the
C-2D \cite{Coniglio1977,janke-clusters-2005} fixed point where $T_p
= T_c$, and in the clean 3D Ising model\cite{sykes-gaunt-1976,
Nagao:1980vw} where $T_p < T_c$.

There are two cases within Eqn.~\ref{eqn:model}  for which $T_p <
T_c$: the clean and  random bond 3D Ising models
\cite{sykes-gaunt-1976,dotsenko-1995,Coniglio1977,berche-janke-2004}.
In these two cases, we claim that as the geometric clusters
percolate  at $T_p$, they display a type of criticality, even though
the system is not at the thermodynamic critical point. To understand
why, it is first helpful to have an intuition about how $T_p$ in  an
interacting model is related to uncorrelated percolation. The
relation of the percolation temperature $T_p$ to  $T_c$ in random
Ising models is constrained by the site percolation thresholds in
the uncorrelated case: on a square lattice, the percolation
threshold is $p_c = 0.59$, while on a cubic lattice, it is $p_c =
0.31,$ where $p$ is the fraction of sites that are occupied.  (In
the Ising case, $p$ becomes the fraction of sites with aligned spins
\cite{Coniglio1977}.)

The high temperature limit of Eqn.~\ref{eqn:model} maps to
uncorrelated percolation with $p = 0.5$. So, in 2D, {\em neither} up
nor down spins percolate at $T\rightarrow \infty,$ since $p(T
\rightarrow \infty) = 0.5 < p_c = 0.59$ on a square lattice. Since a
majority geometric cluster {\em spans} the system in the ordered
phase, it must pass through a percolation point at $T_p = T_c$  if
the transition is continuous. Indeed, it has been proven rigorously
that $T_p = T_c$ in clean 2D Ising models \cite{Coniglio1977}.
Consistent with the argument above, it is known that $\Delta_p =
\Delta_c$ in the 2D random field Ising model \cite{seppala-RF-2D}
where $\Delta$ is the disorder strength. We expect similar physics
to obtain in the random bond Ising model in two dimensions.

In three dimensions, the high temperature limit of Eqn.~\ref{eqn:model}
still maps to uncorrelated site percolation with $p=0.5$,
but  now {\em both}
up and down spins span the system, since $p(T\rightarrow \infty) = 0.5 > p_c = 0.31$
on a cubic lattice,
where $p_c$ is the site percolation threshold.
Upon decreasing temperature,
the majority clusters grow below $T_c$,
but since they were already spanning the system,
the majority clusters do not pass through a percolation point.
However, the minority geometric clusters must pass through a percolation point
as they begin to shrink below $T_c$,
although there is no obvious reason why they should do so at $T_c$.
In fact, in the clean 3D case, it has been rigorously shown that
$T_p < T_c$ \cite{sykes-gaunt-1976}.  This inequality also holds in
the random bond case as well \cite{berche-janke-2004}.

At C-2D, the two sets of critical exponents separately satisfy the
same scaling relations. For example, as discussed in
Refs.~\cite{Coniglio1977} and~\cite{janke-clusters-2005}, the
familiar exponent relations $d-2+\eta = 2\beta/\nu=2(d-d_v)$ are
satisfied by the thermodynamic exponents as $d-2+\eta_c = 2
\beta_c/\nu_c=2(d-d_{v,c})$, and the same exponent relations are
separately satisfied by the connectivity function and the geometric
clusters as $d-2+\eta_p =2 \beta_p/\nu_p=2(d-d_{v,p})$. The
thermodynamic exponents in this case
are\cite{kardar-book,diehl-in-domb} $\eta_c = 0.25$, $\beta_c/\nu_c=
1/8$, and $d_{v,c}=15/8$, where the volume fractal dimension
$d_{v,c}$ is derived from the fractal structure of the FK clusters
at the critical point \cite{janke-clusters-2005}. Inserting these
values into the above scaling relations yields
$2-2+0.25=2*(1/8)=2*(2-15/8)$, and the exponents are clearly
self-consistent. The exponents associated with percolation of the
geometric clusters are $\beta_p/\nu_p= 5/96$ and $d_{v,p}=187/96$
\cite{janke-clusters-2005}, where the volume fractal dimension
$d_{v,p}$ is derived from the fractal structure of the geometric
clusters as they percolate at $T_p = T_c$. Inserting these values
into the exponent relation $2 \beta_p/\nu_p=2(d-d_{v,p})$ yields
$2*(5/96)=2*(2-187/96)$, indicating that the geometric clusters also
satisfy scaling relations at C-2D.

Although we have not found an explicit, direct calculation in the
literature of the exponent $\eta_p$ at C-2D, which in this case
should be derived from the pair connectivity function familiar from
percolation theory, our results based on numerical simulations of
the connectivity function $g_{\rm conn}(r)$ at C-2D indicate that
$d-2+\eta_p=0.104\pm0.002$ as shown in Figs.~\ref{fig:C-2D-cxn},
consistent within error bars with the scaling relation $d-2+\eta_p
=2 \beta_p/\nu_p$ ($0.104 \pm 0.002 \approx 2*(5/96)$). Thus we see
that at C-2D, there are two distinct sets of exponents, and that the
geometric exponents also satisfy the scaling relations of
criticality.
This brings us to our second conjecture:
\vspace{11pt}
\\
{\bf Conjecture \#2:}:  There are two distinct sets of critical exponents which both satisfy scaling relations at {\em all} critical points for which
$T_p = T_c$: one derived from the FK clusters, and the other derived from the geometric clusters.

 \section{Numerical results for the critical exponents defined
 on 2D slices of a 3D system}
 \label{sxn:C-3Dx}
We are ultimately interested in understanding the origin of the
complex, 2-component pattern formation observed at the surfaces of
some strongly correlated electronic
systems\cite{phillabaum-2012,shuo-vo2,Dagotto} via scanning image
probes.   In cases where the data show multiscale clusters of two
distinct types, one of the drivers of such behavior can be proximity
to a critical point of an Ising model, Eqn~\ref{eqn:model}.  We
focus in this paper on the clean Ising case;  future work will
include the effects of disorder. The problem at hand, then, is to
map the properties of the observed geometric clusters to critical
exponents of the model. We focus here on five exponents that that
can  be directly extracted from the microscopy imaging data, and can
be used to compare with different models to identify the
universality class (explained more fully below):  The Fisher
exponent $\tau$ which describes the cluster size distribution; the
volume fractal dimension $d_v$ and the hull fractal dimension $d_h$
of geometric clusters; the anomalous dimension $\eta_c^{||}$ which
controls the spin-spin correlation function; and as we will see
below, $\eta_p^{||}$ which controls the pair connectivity function.

In this paper we keep using Potts model with $q=2$, since a Potts
model with 2 states is an Ising model with half the critical
temperature \cite{RevModPhys.54.235}. We are interested in this
section in the critical behavior displayed by the 2D slice geometric
clusters near the 3D clean Ising phase transition (C-3Dx) which
occurs at the bulk 3D transition temperature $T_c^{3D}$. (In
Sec~\ref{sxn:coniglio} we will demonstrate that this temperature
indeed coincides with the 2D slice percolation point $T_p^{slice}$.)
The simulations of the clean Ising system are done near $T_c^{3D}$
for the study of geometric criticality on 2D slices, and the Wolff
single cluster algorithm \cite{Wolff} is known to do a good job in
mitigating critical slowing down effects near criticality.
Therefore, we use this well-established cluster updating algorithm
in our simulations. We adopt open boundary condition, because this
representation is consistent with our context of microscopy
experiments on real materials with finite scale. Throughout the
paper, we use the logarithmic binning method which is a standard
technique for analyzing the scaling behavior \cite{Newman2005}. For
the 3D cubic lattice, we use $T_c^{3D}=2.25576393$
\cite{C3D_Tc_Livet, C3D_Tc_Talapov},
and all simulations are performed using system sizes from $L^3=80^3$
to $L^3=192^3$ averaged over a large number of configurations in the
Monte Carlo simulation in the order of magnitude $10^4$. We take
samples of configurations every 1000 Wolff steps after the system
reaches equilibrium, in order to mitigate explicit spatial
correlations between different images. Because one whole cluster is
updated for each Wolff iteration, 1000 steps are empirically a large
enough interval to make two neighboring sample images look totally
different.

In the simulations, we have found that the percolation critical
exponents extracted from $(+)$-clusters and $(-)$-clusters
respectively are consistent with each other. This is expected, since
at $T_p^{\rm slice}=T_c^{3D}$ there is $(+)/(-)$ symmetry (see
Section.\ref{sxn:coniglio}).
Therefore we present
the results with both clusters considered together for better
statistics.

The cluster size distribution $D(s)$ measures the number of
geometric clusters with a given size $s$, and scales with
$s^{-\tau}$ as geometric clusters become critical. It is important
to remember that when considered in the bulk context, geometric
clusters do {\em not} display criticality at the bulk transition
temperature $T_c^{3D}$, but rather the minority clusters experience
a percolation transition (and therefore display criticality) at a
lower temperature $T_p^{3D} < T_c^{3D}$ \footnote{At high
temperatures in three dimensions, both $(+)$ and $(-)$ clusters span
the system, since the uncorrelated percolation transition in a cubic
lattice occurs at a fraction of 31$\%$. At very low temperatures in
3D, the system is spanned only by a majority cluster. The
percolation transition for geometric clusters happens inside the
ordered phase, $T_p^{3D} < T_c^{3D}$, when minority clusters cease
to span the system.}. However, we find that when a 2D {\em slice} of
the system is considered, the geometric clusters as defined on the
slice do display critical behavior in the form of power law scaling
at the bulk transition temperature $T_c^{3D}$. For a 2D slice
embedded in the 3D bulk system, we find that $D(s)\sim s^{-\tau}$ at
$T_c^{3D}$, with the Fisher exponent $\tau$ specific for geometric
clusters defined on a 2D cross section of a 3D system. With a finite
field of view (FOV), $D(s)$ in the bulk is known to have a scaling
bump, which skews $\tau$ to a lower value \cite{dahmen-perkovic}.
 In order to mitigate this effect, we analyze
$D(s)$  using only internal clusters, {\em i.e.} those which do not touch a boundary.
$D(s)$ of the internal clusters shows unskewed power law
behavior within a cutoff \cite{sethna-window}.

\begin{figure}
\centering
\includegraphics[width=1.05\columnwidth]{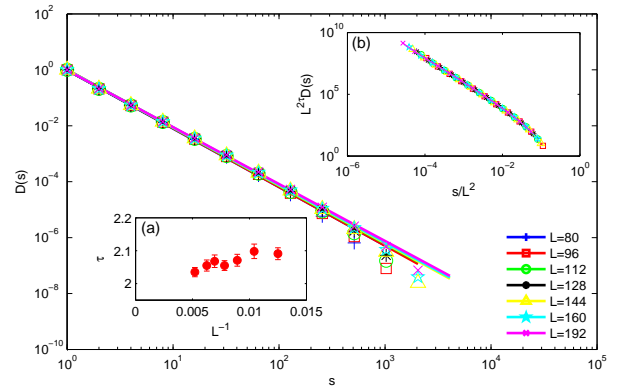}
\caption{Cluster size distribution $D(s)\sim s^{-\tau}$ of internal
C-3Dx geometric clusters at $T_p^{\rm slice}=T_c^{3D}$ for system
sizes from $L=80$ to $L=192$. The inset (a) shows the extrapolation
of $\tau$ from the DLD fits in the main panel to the thermodynamic
limit $L\to \infty$, which gives $\tau=2.001\pm0.013$. The inset (b)
shows the scaling collapse of the curves in the main panel, using
the extrapolated $\tau$ in (a).} \label{tauboth}
\end{figure}

Figure~\ref{tauboth} shows the cluster size distribution $D(s)$ of
internal 2D cross-sectional clusters for different system sizes at
$T_p^{\rm slice}=T_c^{3D}$. For each system size, we exclude the
last two points, which deviate from power law scaling, and fit all
the other points within this cutoff \cite{sethna-window}. Based on
the observed power law behaviors in Fig.~\ref{tauboth}, we use the
discrete logarithmic derivative method (DLD) \cite{Middleton2002} to
extract $\tau$. Since we are only interested in the slope $-\tau$,
we normalize all the $D(s)$ with $D(s=1)=1$ for better comparison
between different system sizes. As can be seen in the figure,
depending on the system size, power law scaling persists for about
2.5 decades of scaling. We extrapolate $\tau$ extracted at
$T_c^{3D}$ to $L\to \infty$, and we find that a linear fit of
$\tau^{(L)}$ yields $\tau=2.001\pm0.013$ as $L \to \infty$.
Here, the extrapolation follows the linear regression method with
error-in-variables \cite{linear_extpl_error}. In general, the
critical exponents cannot be directly obtained from a rigorous
parametric fitting because the exact analytical form of the finite
scaling functions are usually unknown. Under this case, linear
extrapolation is a standard first-order approximation method for
exponents extractions \cite{YL-thesis}, and we use this for critical
exponents extrapolations throughout the whole paper. The natural
finite-size scaling hypothesis for the cluster size distribution at
criticality reads $D(s,L)=(L^2)^{-\tau} \hat{\mathcal{D}}(s/L^2)$,
where $\hat{\mathcal{D}}$ is a universal scaling function
\cite{YL-thesis}.
The inset (b) in Fig.~\ref{tauboth} shows that using the exponent
$\tau$ derived as above, the results from the main panel exhibit
scaling collapse to a universal scaling function $\hat{\mathcal{D}}$
as expected.

\begin{figure}
\centering
\includegraphics[width=1.05\columnwidth]{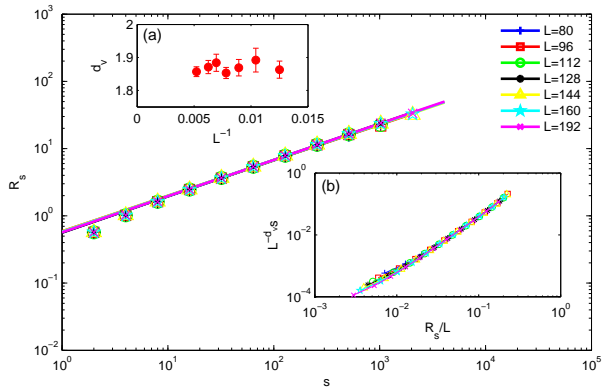}
\caption{The power law fits for $R_s\sim s^{1/d_v}$, using internal
C-3Dx geometric clusters at  $T_p^{\rm slice}=T_c^{3D}$ for system
sizes from $L=80$ to $L=192$. The inset (a) shows the extrapolation
of $d_v$ from the fits in the main panel to the thermodynamic limit
$L\to \infty$, which gives $d_v=1.856\pm0.018$. The inset (b) shows
the scaling collapse of curves in the main panel, using the
extrapolated $d_v$ in (a).} \label{dvboth}
\end{figure}

The cluster volumes $s$ and hulls $h$ become fractal near a
percolation point. The fractal nature of  the 2D cross-sectional
cluster volumes at $T_p^{\rm slice}=T_c^{3D}$ can be described by $s
\sim R_s^{d_v}$ with $s\gg1$.  Here, $d_v$ is the volume fractal
dimension, and $R_s$ is the radius of gyration of the cluster,
defined by $2R_s^2=\sum_{i,j}|{\bf r}_i-{\bf r}_j|^2/s^2$, where the
sum is over sites in the cluster, and ${\bf r}_i$ and ${\bf r}_j$
are the positions of the $i$th and $j$th sites \cite{Percobook}.
Figure~\ref{dvboth} shows a DLD power law fit of $s \sim R_s^{d_v}$,
using only the internal clusters to mitigate boundary effects. Since
the scaling relation is for $s\gg1$, we exclude the first 3 points
which correspond to the non-universal short length regime. All the
other points belong to the power law scaling regime. The power law
fits at $T_p^{\rm slice}=T_c^{3D}$ persist for
about two decades of scaling, depending on the system size. A
straightforward fit extrapolating $d_v$ to the thermodynamic limit
yields $d_{v}=1.856\pm0.018$. The inset Fig.~\ref{dvboth}(b)
illustrates the scaling collapse for different system sizes to the
universal function $\hat{s}$ using the extrapolated values of $d_v$,
based on the finite-size scaling hypothesis $s=L^{d_v}
\hat{s}(R_s/L)$ \cite{YL-thesis}.

\begin{figure}
\centering
\includegraphics[width=1.05\columnwidth]{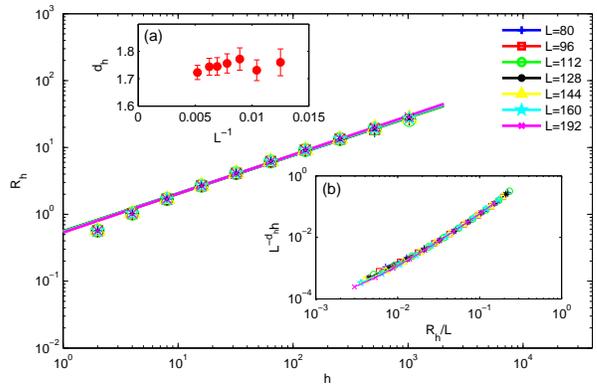}
\caption{The power law fits for $R_h\sim h^{1/d_h}$, using internal
C-3Dx geometric clusters at $T_p^{\rm slice}=T_c^{3D}$ for system
sizes from $L=80$ to $L=192$. The inset (a) shows the extrapolation
of $d_h$ from the fits in the main panel to the thermodynamic limit
$L\to \infty$, which gives $d_{h}=1.714\pm0.022$. The inset (b)
shows the scaling collapse of curves in the main panel, using the
extrapolated $d_h$ in (a)} \label{dhboth}
\end{figure}

The cluster hulls $h$ scale with
$h \sim R_h^{d_h}$ near a percolation point. In this case, we are
only tracking the outer (externally accessible) surfaces of each
cluster. Thus one surface might contain not only the cluster itself,
but also the subclusters inside. Therefore, $R_h$ here refers to the
radius of gyration of all the sites enclosed by the hull, including
any subclusters. To mitigate the boundary effect, we still use only
the internal clusters in the DLD fit. We exclude the first 2 point
from the fit ($h$ is a smaller measure than $s$ for a cluster),
which deviate from the power law scaling regime and correspond to
non-universal short distance physics. As shown in Fig.~\ref{dhboth},
the power law behavior extends over
about 2 decades, depending on the system size. A straightforward fit
extrapolating to the thermodynamic limit at $T_p^{\rm
slice}=T_c^{3D}$ yields $d_{h}=1.714\pm0.022$. The inset
Fig.~\ref{dhboth}(b) shows the scaling collapse for different system
sizes to the universal function $\hat{h}$ using the extrapolated
values of $d_h$, based on the finite-size scaling hypothesis $h=
L^{d_h} \hat{h}(R_h/L)$ \cite{YL-thesis}.

\begin{figure}
\centering
\includegraphics[width=1.05\columnwidth]{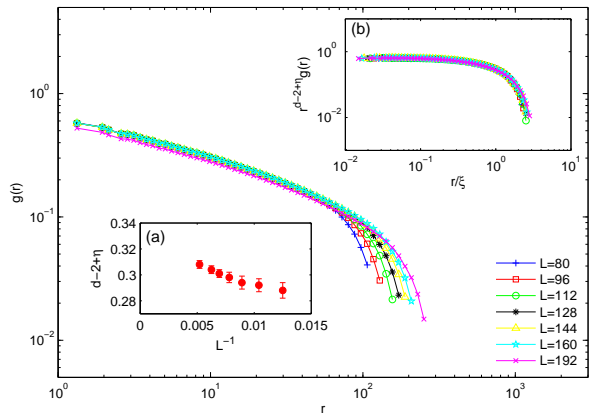}
\caption{Pair connectivity function $g_{\rm conn}(r)$ on the 2D
slice of clean 3D Ising system with different system sizes from
$L=80$ to $L=192$ at $T_p^{\rm slice}=T_c^{3D}$. The inset (a) shows
the extrapolation of $d-2+\eta_p$ from the scaling form fits of
$g_{\rm conn}(r)$ in the main panel to the thermodynamic limit $L\to
\infty$, which yields $d-2+\eta_p=0.322\pm0.002$. The inset (b)
shows the scaling collapse of curves in the main panel, using the
extrapolated $d-2+\eta_p$ in (a).} \label{fig:C-3Dx-cxn}
\end{figure}

As introduced in the previous section, the pair connectivity
function $g_{\rm conn}(r)$ is defined as the probability that two
sites separated by a distance $r$ belong to the same finite cluster.
This correlation function scales with
$g_{\rm conn}(r)\sim r^{-(d-2+\eta_p)}$ close to the percolation
point.
Fig.~\ref{fig:C-3Dx-cxn} shows $g_{\rm conn}(r)$ at $T_p^{\rm
slice}=T_c^{3D}$ with different system sizes. As shown by the
figure, in addition to the power law behavior, associated with
$g_{\rm conn}(r)$ there appears to be an exponential decay
characterized by the correlation length scale of the finite system.
So we infer the scaling form $g_{\rm conn}(r)\sim
r^{-(d-2+\eta_p)}e^{-r/\xi_p}$ with a finite correlation length, and
then fit the curves in the main panel using this scaling form. From
the fits, we find that the curves of $g_{\rm conn}(r)$ is consistent
with the shape of scaling form with some large $\xi_p$ comparable
with the finite system size. The inset (a) of
Fig.~\ref{fig:C-3Dx-cxn} shows the extracted anomalous dimension
$d-2+\eta_p$ from the scaling form fits in the main panel. A
straightforward fit extrapolating to $L\to \infty$ yields
$d-2+\eta_p=0.322\pm0.002$.
The curves in the main panel collapse onto each other for the
extrapolated $d-2+\eta_p$, as shown by the inset (b) of
Fig.~\ref{fig:C-3Dx-cxn}.

\begin{figure}
\centering
\includegraphics[width=1.05\columnwidth]{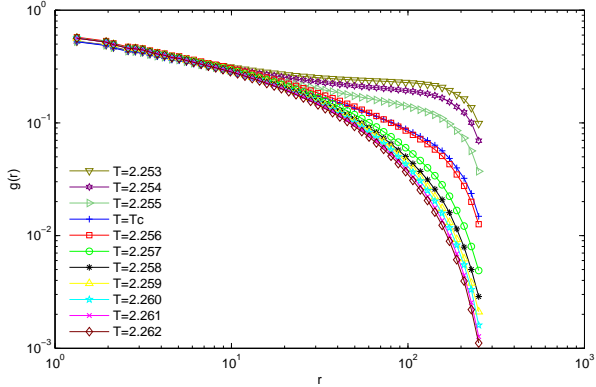}
\caption{Pair connectivity function $g_{\rm conn}(r)$ on the 2D
slice of the clean 3D Ising system with size $L=192$ at different
temperatures around $T_p^{\rm slice}=T_c^{3D}$. $g_{\rm conn}(r)$
turns up at large $r$ for $T< T_c^{3D}$.} \label{fig:C-3Dx-cxn-T}
\end{figure}

In Fig. \ref{fig:C-3Dx-cxn-T}, we show the C-3Dx pair connectivity
function $g_{\rm conn}(r)$ simulated with system size $L=192$ for
different temperatures near $T_c^{3D}$.  It is evident that for
$T\gtrsim T_c^{3D}$, $g_{\rm conn}(r)$ is consistent with the shape
of the scaling form $g_{\rm conn}(r)\sim
r^{-(d-2+\eta_p)}e^{-r/\xi_p}$; and for $T< T_c^{3D}$, $g_{\rm
conn}(r)$ turns up at large r. For $g_{\rm conn}(r)$ calculated
using all the possible pairs in the system, this up-turn is a
characteristic behavior of the connectivity function when the system
transitions from the ``disordered side'' (zero infinite network
strength $P=0$) to the ``ordered side'' (non zero infinite network
strength $P>0$) through a percolation point, indicating the presence
of an infinite cluster on a 2D slice. (Here $P$ is the percolation
order parameter defined on the 2D slice.) Therefore, this behavior
can be used as a {\em diagnostic tool} to estimate the percolation
point in an experimental situation. In the present case, this
up-turn happens in the vicinity of $T_c^{3D}$, corroborating the
idea that $T_p^{\rm slice}$ coincides with the bulk thermodynamic
transition temperature, $T_p^{\rm slice}=T_c^{3D}$.

We have studied the scaling behavior of cross-sectional 2D slice
geometric clusters embedded in the 3D bulk clean Ising system. We
have found robust power law scaling (about two decades) for the
measures related to the self-similarity of geometric clusters at
$T_c^{3D}$, as well as the existence of universal scaling functions.
All of these phenomena corroborate the conjecture that geometric
criticality on a 2D slice occurs at the bulk critical temperature
$T_p^{\rm slice}=T_c^{3D}$, and we have extracted the corresponding
critical cluster exponents, as summarized in Table~\ref{c2n3d}.

\section{Numerical Results for the Connectivity Function
in the Clean 2D Ising Model}
\label{sxn:C-2D}

\begin{figure}
\centering
\includegraphics[width=1.05\columnwidth]{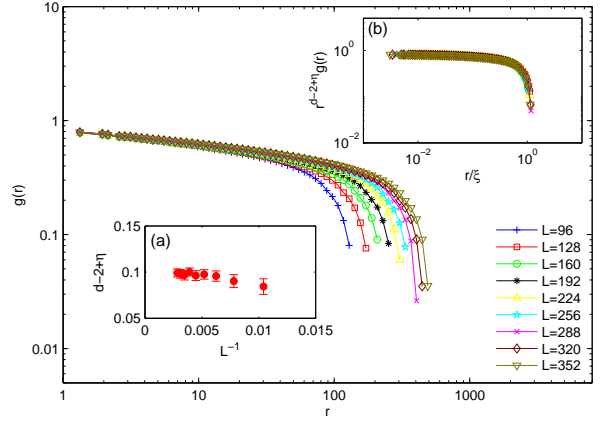}
\caption{Pair connectivity function $g_{\rm conn}(r)$ of the 2D
clean Ising system with different sizes from $L=96$ to $L=352$ at
$T_p^{2D}=T_c^{2D}$. The inset (a) shows the extrapolation of
$d-2+\eta_p$ from the scaling form fits of $g_{\rm conn}(r)$ in the
main panel to the thermodynamic limit $L\to \infty$, which yields
$d-2+\eta_p=0.104\pm0.002$. The inset (b) shows the scaling collapse
of curves in the main panel, using the extrapolated $d-2+\eta_p$ in
(a)} \label{fig:C-2D-cxn}
\end{figure}

\begin{figure}
\centering
\includegraphics[width=1.05\columnwidth]{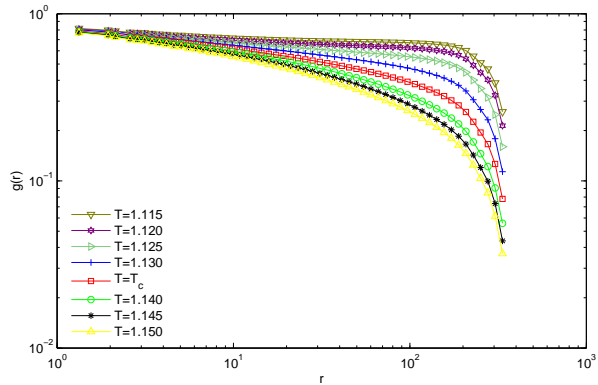}
\caption{Pair connectivity function $g_{\rm conn}(r)$ of the 2D
clean Ising system with size $L=256$ at different temperatures
around $T_p^{2D}=T_c^{2D}$. The change of behavior for $g_{\rm
conn}(r)$ from the scaling form shape to the turning-up shape
happens in the close vicinity of $T_p^{2D}=T_c^{2D}$}
\label{fig:C-2D-cxn-T}
\end{figure}

In the clean 2D Ising model, the percolation point coincides with
the critical point, and both up and down geometric clusters
percolate symmetrically at $T_p^{2D}=T_c^{2D}$ under zero external
field \cite{Coniglio1976,Coniglio1977}. The values of the exponents
$\tau$, $d_v$ and $d_h$ are already given in the work by Janke et.
al. \cite{janke-clusters-2005}, as summarized in Table~\ref{c2n3d}.
In the present work, we simulate the
 connectivity
function in order to obtain the value of the anomalous dimension
$d-2+\eta_p$. Fig. \ref{fig:C-2D-cxn} shows the connectivity
function $g_{\rm conn}(r)$ at $T_p^{2D}=T_c^{2D}$, with
$T_c^{2D}=1/ln(1+\sqrt{2})$ for the square lattice
\cite{RevModPhys.54.235}. The inset (a) shows $d-2+\eta_p$ extracted
from the scaling form fit of the connectivity function $g_{\rm
conn}(r)\sim r^{-(d-2+\eta_p)}e^{-r/\xi_p}$ at $T_c^{2D}$ for
different system sizes, and a straightforward fit extrapolating to
the thermodynamic limit gives $d-2+\eta_p=0.104\pm0.002$. With this
value, all of the curves in the main panel collapse on top of each
other, as shown by the inset (b). Figure ~\ref{fig:C-2D-cxn-T} shows
the behavior of $g_{\rm conn}(r)$ around $T_c^{2D}$ with system size
fixed at $L=256$. As for the case of C-3Dx, we find that for
$T>T_c^{2D}$, the pair connectivity function $g_{\rm conn}(r)$ is
consistent with the scaling form shape and for $T<T_c^{2D}$ $g_{\rm
conn}(r)$ turns up. This change of behavior happens in the close
vicinity of $T_c^{2D}$, consistent with the idea that
$T_p^{2D}=T_c^{2D}$.

\section{Relation of infinite cluster to bulk magnetization on a 2D slice}
\label{sxn:coniglio}

\begin{figure}
\centering
\includegraphics[width=0.95\columnwidth]{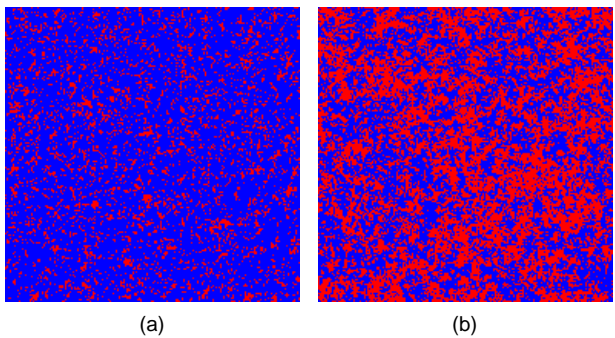}
\caption{Illustration of the equilibrium 2D slice configuration
embedded in the 3D Ising system with L=192 at (a)$T=T_p^{3D}$ and
(b)$T=T_c^{3D}$. } \label{C3Dx_config}
\end{figure}

The bulk minority geometric clusters for the clean 3D Ising model
are known to percolate inside the ordered phase, with the
percolation temperature $T_p^{3D}=0.93T_c^{3D}<T_c^{3D}$
\cite{sykes-gaunt-1976,Nagao:1980vw}. Therefore the bulk geometric
clusters {\em do not}  display fractal properties at the critical
temperature $T_c$, although they do at $T_p^{3D}$. At first glance,
one might think that the geometric clusters on a 2D slice, which are
cross-sections of the 3D bulk clusters, should also display critical
scaling behavior at the bulk percolation temperature $T_p^{3D}$.
However, as shown by Fig. \ref{C3Dx_config}(a), the geometric
clusters defined on a 2D cross-section  at $T_p^{3D}$ do not display
self-similar scaling behavior, since the minority clusters are
small, and a single majority cluster spans the slice. This shows
that $T_p^{3D}$ is not the percolation point for 2D slice geometric
clusters. On the other hand, inspired by the fact that the
percolation point coincides with the critical point for 2D Ising
model \cite{Coniglio1977}, we find that at $T_c^{3D}$, as shown by
Fig. \ref{C3Dx_config}(b), the 2D cross-sectional geometric clusters
indeed show self-similarity and fractal behavior over all
lengthscales in the field of view. This is consistent with the
findings of Ref.~\onlinecite{saberi-C3Dx}, which argues that
$T_p^{\rm slice}=T_c^{3D}$. Also, our numerical simulation in
Section~\ref{sxn:C-3Dx} corroborates $T_p^{\rm slice}=T_c^{3D}$ from
various perspectives, including robust power law behaviors of 2D
slice geometric clusters at $T_c^{3D}$ and change of behavior of
pair-connectivity function through $T_c^{3D}$.

In the clean 2D Ising model, Coniglio and co-workers
\cite{Coniglio1977} showed that the percolation temperature
coincides with the thermodynamic transition temperature, $T_p =
T_c$, by showing that $M < R$, where  $R$ is the infinite network
strength,
and $M$ is the net magnetization. This comes from the very physical
idea that in a 2D system with thermal fluctuations, a net
magnetization requires that at least one geometric cluster spans the
system. From Fig.~\ref{C3Dx_config}, it appears that something
similar must be going on in the 3D Ising model, when clusters are
defined an a 2D slice. As will be discussed later in this section,
our numerical results indeed show that $M < R^{slice}$, where
$R^{slice}$ is the infinite network strength defined on a plane of a
3D system \footnote{Note that in Ref.~\onlinecite{Coniglio1976},
Coniglio {\em et al.} showed that $M < R$ in any dimension, for $T <
Tc$. The question here is whether $M < R_x$ on a 2D slice in a 3D
system.}.

To rigorously prove  that $T_p^{\rm slice}=T_c^{3D}$ would require a
proof the following lemma: For a 2D slice of the 3D ferromagnetic
Ising model with nearest-neighbor interactions at zero external
field ($H=0$) and below the critical temperature($T<T_c$), we have
\begin{eqnarray}
M(0^+,T)\le R^{slice}_\uparrow(0^+,T) - R^{slice}_\downarrow(0^+,T).
\end{eqnarray}
The symbol $0^+$($0^-$) means $H=0$ with $(+)$-boundary
($(-)$-boundary) condition. For our discussion, we limit ourselves
to the $(+)$-boundary by convention, and the results of
$(-)$-boundary follow those of $(+)$-boundary under symmetry.
$M(0^+,T)$ is the reduced spontaneous magnetization on the 2D slice
(which is identical to the reduced spontaneous magnetization of the
bulk system at thermodynamic limit), and is given by
$M(0^+,T)=\langle\tilde{\pi}_0\rangle_+-\langle\pi_0\rangle_+$. Here
$\pi_i=\frac{1}{2}(1-\sigma_i)$ and
$\tilde{\pi}_i=\frac{1}{2}(1+\sigma_i)$ are the lattice gas
variables relative to the $i$th site, and $\sigma_i=\pm 1$ is the
usual Ising spin variable. $\tilde{\pi}_i$ ($\pi_i$) is equivalently
the characteristic function of the event that spin in $i$ is up
(down). The characterstic function of an event is that when the
event happens, the function takes the value 1, otherwise it takes
the value 0. The subscript $0$ refers to the origin, and
$\langle\rangle_+$ means the thermal average for the Ising model
with $(+)$-boundary. Here we only take one slice for each Ising
configuration to be counted in the thermal average, and we do not
average over slices belonging to the same configuration in order to
reduce spatial dependence. Therefore there is a one-to-one
correspondence between the slice configuration and bulk system
configuration, and by convention we fix the slice to be the $z=0$
plane in the $\mathbb{Z}^3$ cubic lattice so that the slice contains
the origin.
$R^{slice}_\uparrow(0^+,T)$($R^{slice}_\downarrow(0^+,T)$) is the 2D
slice infinite network strength for up(down) spins, defined as the
probability that the origin belongs to the infinite 2D slice
$(+)$-cluster ($(-)$-cluster), or equivalently it is the weight of
the infinite 2D slice $(+)$-cluster($(-)$-cluster). Mathematically,
it is given by
$R^{slice}_\uparrow(0^+,T)=\langle\tilde{\gamma}_0^\infty\rangle_+$
($R^{slice}_\downarrow(0^+,T)=\langle\gamma_0^\infty\rangle_+$),
where $\tilde{\gamma}_i^\infty$($\gamma_i^\infty$) is the
characteristic function of the event that spin in $i$ belongs to the
infinite 2D slice $(+)$-cluster($(-)$-cluster).

For the 2D slice of a cubic lattice system, which is a simple planar
graph admitting an elementary cell and two axes of symmetry, we have
the statement that (i) an infinite 2D slice $(+)$-cluster cannot
coexist with an infinite 2D slice $(-)$-cluster, and (ii) the number
of each type of infinite 2D slice cluster must be either 0 or 1
\cite{Coniglio1977, Coniglio1976, Harris1960, Fisher1961,
Newman1981}. This means that
\begin{eqnarray}
R^{slice}_\uparrow(H,T)R^{slice}_\downarrow(H,T)=0,
\end{eqnarray}
and thus there can only be at most one infinite $(+)$-cluster on the
2D slice with the $(+)$ boundary condition. Therefore, Eqn.~(2)
equivalently becomes
\begin{eqnarray}
M(0^+,T)\le R^{slice}_\uparrow(0^+,T),
\end{eqnarray}
as intuitively discussed in the previous paragraph.

\begin{figure}
\centering
\includegraphics[width=0.95\columnwidth]{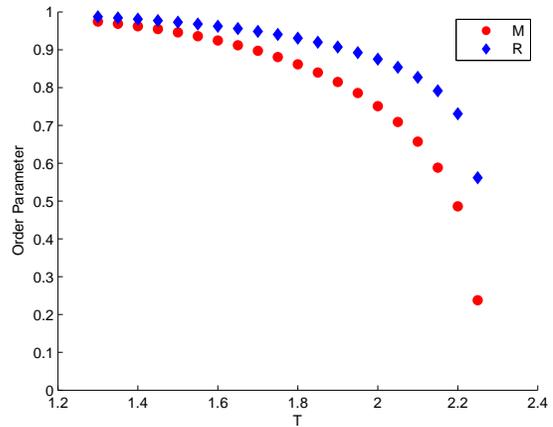}
\caption{Numerical results of the reduced spontaneous magnetization
M and the infinite network strength R of the 2D slice embedded in 3D
Ising system under zero external field with $L=192$ at $T<
T_c^{3D}$, averaged over a number of configurations in the order of
magnitude $\sim10^5$. M and R both approach 1 when T decreases, with
$M < R$.} \label{C3Dx_MR}
\end{figure}

The rigorous demonstration of (4) requires that both the Markov
property \cite{markov}
and the Fortuin-Kasteleyn-Ginibre (FKG) inequality \cite{FKG} both
hold \cite{Coniglio1977}. The FKG inequality \footnote{ The FKG
inequality concerns the relation of the expectation value of the
product of two functions of Ising variables to the product of the
expectation values. See Refs.~\onlinecite{FKG} and
\onlinecite{Coniglio1977}.} may be directly extended to the 2D slice
since it is applicable to any finite set in the bulk system. However
the Markov property cannot be restricted to a 2D slice, since the
Markov chain may extend off of the slice in a 3D system.
Rather than delving into a mathematical physics approach, for the
purpose of this paper, we numerically demonstrate the fact that the
inequality (4) holds for the 2D slice embedded in the 3D Ising
model, as revealed by Fig. \ref{C3Dx_MR}.

By symmetry we have $R^{slice}_\uparrow(H,T)=R^{slice}_\downarrow(-H,T)$.
Therefore from Eqn.~(3),
for $T\geq T_c^{3D}$, we must have
$R^{slice}_\uparrow(0,T)=R^{slice}_\downarrow(0,T)=0$.
For $T<T_c^{3D}$, Eqns.~(3) and (4)
yield $R^{slice}_\downarrow(0^+,T)=0$ and $R^{slice}_\uparrow(0^+,T)\geq
M(0^+,T)>0$.
If we suppose that the finite cluster weight is
continuous and has a limit at $T_c^{3D}$, then due to symmetry we
have $\lim_{T\to T_c^{3D}}
\langle\tilde{\gamma}_0\rangle_+=\lim_{T\to T_c^{3D}}
\langle{\gamma}_0\rangle_+$.  From
\begin{eqnarray}
M(0^+,T)&=&\langle\tilde{\pi}_0\rangle_+-\langle{\pi}_0\rangle_+ \nonumber \\
&=&\langle\tilde{\gamma}_0\rangle_+-\langle{\gamma}_0\rangle_+
+\langle\tilde{\gamma}_0^\infty\rangle_+-\langle{\gamma}_0^\infty\rangle_+,
\end{eqnarray}
we have $\lim_{T\to T_c^{3D}}
\langle\tilde{\gamma}_0^\infty\rangle_+=\lim_{T\to
T_c^{3D}}\langle{\gamma}_0^{\infty}\rangle_+$, which then leads to
$\lim_{T\to T_c^{3D}} R^{slice}_\uparrow(0^+,T)=0$, so that no jump
occurs at $T_c^{3D}$. This is intuitively consistent with a
continuous phase transition. In other words, under the $(+)$
boundary condition with zero external field, the 2D slice
$(+)$-cluster percolates at $T_c^{3D}$, as characterized by the
percolation order parameter $R^{slice}_\uparrow(0^+,T)$ which
continuously changes from 0 to non-zero.

All of the above statements apply to the $(-)$-cluster with $(-)$ boundary
condition under symmetry. Thus, for zero applied field $H=0$, the $(+)$-cluster and
$(-)$-cluster defined on a 2D slice both approach the percolation point at the limit
$T\rightarrow T_c^{3D}$. The spontaneous infinite cluster type in
the ordered phase $T < T_c^{3D}$ is decided by the boundary
condition (whether $(+)$ or $(-)$), and is symmetric under $H=0^+
\leftrightarrow H=0^-$. Therefore with $H=0$ at the limit
$T\rightarrow T_c^{3D}$, both types of clusters percolate on the slice, and due
to symmetry they present the same universal scaling behavior with
the same critical exponents.

As supported by our numerical simulation as summarized in
Table~\ref{c2n3d}, this critical percolation point of geometric
clusters defined on 2D slices of the bulk system in the 3D clean
Ising model is a {\em new universality class} of geometric
criticality, distinct from that of the clean 2D model and
uncorrelated percolation in 2D, as well as that of the bulk
percolation which occurs inside the ordered phase of the 3D system,
since $T_p^{3D}\neq T_p^{\rm slice}$.

\section{Two correlation lengths}
\label{sxn:cl}
\begin{figure}
\centering
\includegraphics[width=0.95\columnwidth]{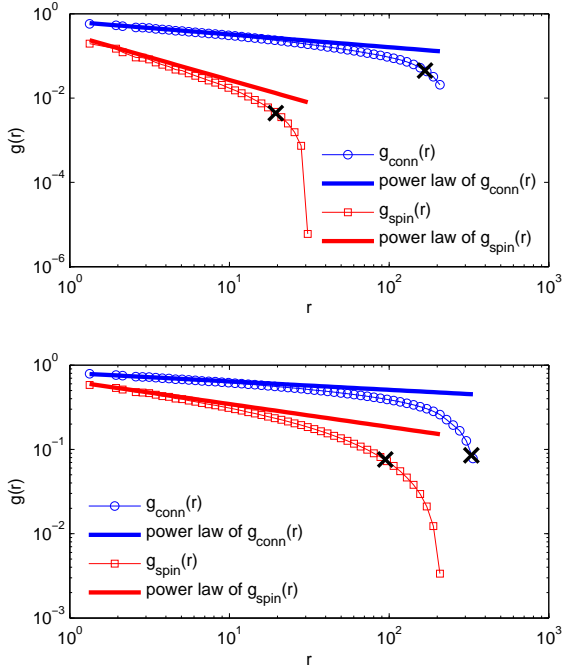}
\caption{Pair connectivity function $g_{\rm conn}(r)$ and spin-spin
correlation function $g_{\rm spin}(r)$ for (a) C-3Dx with $L=160$ at
the critical temperature $T_c^{3D}(L \rightarrow \infty)$ and (b)
C-2D with $L=256$ at the critical temperature $T_c^{2D}(L
\rightarrow \infty)$. The black cross markers on the correlation
functions denote the lengthscales of correlation lengths. The thick
lines represent the power law behaviors of the correlation functions
by setting the exponential terms to unity}
\label{fig:correlation-length}
\end{figure}

\begin{figure}
\centering
\includegraphics[width=0.95\columnwidth]{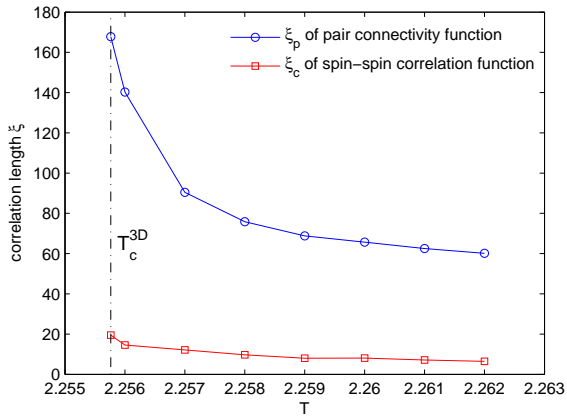}
\caption{The two correlation lengths $\xi_p$ (pertaining to $g_{\rm
conn}(r)$) and $\xi_c$ (pertaining to $g_{\rm spin}(r)$) for C-3Dx
with $L=160$ at $T \gtrsim T_c^{3D}(L\rightarrow \infty)$, computed
by using the scaling form fits of the correlation functions. }
\label{fig:xi_aroundTc}
\end{figure}

In this section we elaborate on the two different correlation
lengths associated with the pair connectivity function and the
spin-spin correlation function. In Fig.~\ref{fig:correlation-length}
we illustrate the pair connectivity function $g_{\rm conn}$ and
spin-spin correlation function $g_{\rm spin}$ for both the 2D slice
of the 3D clean Ising model and 2D clean Ising model at their
respective transition temperatures with specific system sizes
($L=160$ for C-3Dx and $L=256$ for C-2D). Since $g_{\rm spin}$
approaches $0$ for large $r$, it can fluctuate into negative values
under our simulation with finite system sizes and finite number of
configurations for averaging. Once negative values appear, in
Fig.~\ref{fig:correlation-length} we cut off the remaining points
with larger lengthscales on the log-log plots.

By doing scaling form fits of the correlation functions, we can
extract the correlation length for each function. We refer to the
correlation length derived from geometric criticality of the {\em
percolation} order parameter $\xi_p$, extracted from the pair
connectivity function. The standard correlation length used for
thermodynamic criticality of the spin-spin correlation function we
refer to as $\xi_c$. In Fig.~\ref{fig:correlation-length}, which is
calculated at the thermodynamic critical temperature $T_c^{3D}(L
\rightarrow \infty)$ of the infinite size system for C-3Dx, these
two correlation lengths are $\xi_p=167.81$ and $\xi_c=19.50$.
At the thermodynamic critical temperature $T_c^{2D}(L \rightarrow
\infty)$ of the infinite size system for C-2D, these correlation
lengths are $\xi_p=327.38$ and $\xi_c=94.65$. The fact that the
correlation lengths are not infinite at this point is due to the
fact that the simulated systems are finite and that the
corresponding critical temperatures are nicely shifted from its
infinite value because of finite size effects.

Notice that in both cases, $\xi_p$ is much larger than $\xi_c$. From
the scaling form fits, we have $d-2+\eta_c=1.078\pm0.068$ for C-3Dx,
and $d-2+\eta_c=0.272\pm0.008$. (The results for $d-2+\eta_p$ from
the fits are already reflected in the insets (a) of
Fig.~\ref{fig:C-3Dx-cxn} and Fig.~\ref{fig:C-2D-cxn}) They are very
close to the established values $d-2+\eta_c=1.0336$ (3D clean Ising
model) \cite{C3Dtau} and $d-2+\eta_c=0.25$ (2D clean Ising model)
\cite{C2Dspin1,kardar-book}, corroborating the validity of our
results for the correlation lengths. As implied in our previous
discussion, these two different correlation lengths correspond to
two categories of criticality, geometric criticality and
thermodynamic criticality, with two distinct order parameters (the
infinite network strength $R$ and the magnetization $M$
respectively).

For C-2D, it is well established that the correlation length
exponents $\nu_p$ (based on geometric clusters) and $\nu_c$ (based
on FK clusters) are different and are related by $\nu_p \geq \nu_c$
\cite{Coniglio1977}. Therefore, near the critical point
$T_p^{2D}=T_c^{2D}$, from the relation $\xi \sim |T-T_c|^{-\nu}$,
the connectivity correlation length of geometric clusters is
expected to be large compared to the spin-spin correlation length
($\xi_p > \xi_c$). This is consistent with our simulation
Fig.~\ref{fig:correlation-length}(b). Our simulation
Fig.~\ref{fig:correlation-length}(a) reveals that $\xi_p > \xi_c$
{\em also} applies for C-3Dx.
To further validate this inequality, in Fig.~\ref{fig:xi_aroundTc},
we compute the two correlation lengths for C-3Dx geometric
criticality as a function of temperature near the critical point
$T_p^{slice}=T_c^{3D}$ on the disordered side, where the correlation
function is consistent with the scaling form shape.
 As shown by Fig.~\ref{fig:xi_aroundTc}, $\xi_p$ is
always much larger than $\xi_c$, and they increase as $T \rightarrow
T_c$. This suggests that the exponent inequality $\nu_p \geq \nu_c$
also applies for C-3Dx.

\section{Discussion}
\label{sxn:discussion}
\begin{table*}[ht]
\caption{Theoretical/Numerical values for the 2D geometric cluster
self-similarity characterized critical percolation exponents for
Ising models and standard percolation model.} %
\centering %
\begin{ruledtabular}
\begin{tabular}{c c c c c} %
Model & 2D Ising Model \cite{janke-clusters-2005,cardy-book} & 2D slice of 3D Ising model (this work) & Standard Percolation \cite{Stauffer, P2Ddh, Percobook}\\ [0.5ex] %
\hline %
Fixed Point & C-2D & C-3Dx & P-2D & \\
\hline %
$\tau$ & $379/187=2.027$ & $2.001\pm0.013$ & $187/91=2.055$\\ %
$d_v$ & $187/96=1.948$ & $1.856\pm0.018$ & $91/48=1.896$\\
$d_h$ & $11/8=1.375$ & $1.714\pm 0.022$ & $7/4=1.75$\\
$d-2+\eta_p$ & $0.104\pm 0.002$ (this work) & $0.322\pm0.002$ & $5/24=0.208$\\
\end{tabular}
\end{ruledtabular}
\label{c2n3d}
\end{table*}

So far, we have argued that the bulk thermodynamic critical point of
the 3D clean Ising model coincides with the percolation of geometric
clusters defined on a 2D slice, $T_p^{slice} = T_c^{3D}$. Also we
have numerically investigate this geometric criticality by
extracting the critical cluster exponents and studying the behaviors
of the pair-connectivity functions (together with the 2D clean Ising
model). These results have several important implications:

(1)  There are {\em two length scales} in the clean 3D Ising model.
Aside from the usual length scale associated with the spin-spin
correlation length $\xi_c$, we have defined a new length scale,
$\xi_p$ which is the correlation length of the pair connectivity
function of geometric clusters defined on a 2D slice.  We have
furthermore shown that these two length scales diverge differently
as $T \rightarrow T_c^{3D}$, and that the generic case is that
$\xi_p > \xi_c$. One consequence of this is that near this critical
point, experiments which can detect clusters on a slice (for
example, scanning probes like scanning tunneling microscopy) should
reveal that the pair connectivity function defined on a slice is
power law over a larger region of the phase diagram than the
spin-spin correlation function. That is, the critical region appears
to be larger when measured via a pair correlation function than via
a spin-spin correlation function. The argument above also applies to
2D clean Ising model.

(2)  %
As discussed in the context of Conjecture \#2 in
Section~\ref{sxn:exponents}, for the 2D clean Ising model, the
geometric cluster percolation exponents and thermodynamic critical
exponents are different in definition and values, but they satisfy
the same scaling relation in their own closed sets respectively. As
in the purely 2D case, there are also two distinct sets of critical
exponents at $T_c^{3D}$ for the clean 3D Ising model. One set comes
from the geometric clusters defined on a 2D slice of the clean 3D
Ising system, and the other from the bulk FK clusters. Using the
numerical values derived in this work, the scaling relation
$d-2+\eta=2(d-d_v)$ is satisfied within error bars. Inserting the
values from Table~\ref{c2n3d}, $d-2+\eta_p = 0.322 \pm .002$, and
$d_{v,p} =1.856\pm .018$ gives $2(d-d_{v,p}) = 0.288\pm .036~$. The
scaling relation $\tau = (d + d_v)/d_v$ is {\em almost} satisfied:
From the Table, we have  $\tau = 2.001 \pm .013$ and $d_{v,p} =
1.856 \pm .018$, so that  $(d+d_{v,p})/d_{v,p} = 2.078 \pm .010~$.
Recall that the scaling function associated with the exponent $\tau$
suffers from a significant scaling bump, which skews the value of
$\tau$ to lower values in finite size systems. The slight mismatch
in this scaling relation is likely due to this well-known finite
size effect for $\tau$ \cite{dahmen-perkovic}.

We find some indication that the Coniglio inequalities for these two
different sets of critical exponents also hold on a 2D slice of the
3D clean Ising model.  For example, our results in
Fig.~\ref{fig:xi_aroundTc} suggest that the exponent inequality
derived by Coniglio and coworkers \cite{Coniglio1977} for the
correlation length exponents $\nu_p$ and $\nu_c$ at the C-2D
critical point also holds at the C-3Dx critical point, $\nu_p \ge
\nu_c$.

(3) The C-3Dx fixed point represents a {\em new universality class}:
not only are the critical exponents numerically distinct from the
geometric criticality of the 2D clean Ising system and from 2D
uncorrelated percolation, but they are also distinct from the values
of bulk percolation in the 3D clean Ising model, which happens
inside the ordered phase $T_p^{3D}<T_c^{3D}=T_p^{\rm slice}$.

(4)  All of the above implies that there are {\em two order
parameters on the 2D slice of the 3D clean Ising model}, just as
there are two order parameters in the 2D clean Ising model. One of
it is associated with the thermodynamic phase transition, and the
other is associated with the geometric cluster percolation.

\section{Conclusions and Outlook}
\label{sxn:conclusions}
In summary, we have studied  geometric criticality associated with
the correlated percolation of interacting geometric clusters
on a 2D slice of clean 3D Ising models (C-3Dx).
We find that as in the clean 2D Ising model (C-2D),
the geometric criticality associated with
the percolation of interacting geometric clusters at the
C-3Dx critical fixed point corresponds to a unique
{\em geometric universality class}
which is distinct from that of uncorrelated percolation.

In addition, we find that at both the C-3Dx fixed point and the C-2D fixed point,
the geometric clusters become fractal, leading to
power law behavior in the pair connectivity function.
We find that in the vicinity of thermodynamic criticality,
the pair connectivity function displays power law behavior
over a wider region of the phase diagram than
does the spin-spin correlation function.

One consequence of this finding is that
 the pair connectivity function can
be a useful tool for diagnosing criticality in the context of image
data from probes which can take ``slice'' data, and also potentially
from scanning surface image probes such as scanning tunneling
microscopy, atomic force microscopy, and scanning infrared
microscopy. In a future publication, we will explore the relation of
these concepts to surface criticality, which is relevant for the
application of these ideas to scanning surface probes. Especially in
light of the tremendous increase in data coming from a growing
number of scanning image probes \cite{basov-rmp}, it is useful to
have another method of analysis in hand. We expect that future
studies regarding geometric criticality in random bond and random
field Ising models will further facilitate the application of
geometric cluster analyses to the interpretation of the many types
of 2D image probe experimental techniques, especially in cases where
complex pattern formation is observed
\cite{phillabaum-2012,superstripes-erice-2014,shuo-vo2,hoffman-clusters}.

\begin{acknowledgments}
We thank J.~E.~Hoffman, E.~J.~Main, B.~Phillabaum, and C.-L.~Song
for helpful conversations. S. L. and E.~W.~C acknowledge support
from NSF Grant No. DMR-1508236 and Department of Education Grant No.
P116F140459. K.A.D. acknowledges support from NSF Grant No. DMS
1069224 and NSF Grant No. CBET-1336634. This research was supported
in part through computational resources provided by Information
Technology at Purdue, West Lafayette, Indiana.
\end{acknowledgments}

\bibliography{2dslice_ref}

\end{document}